\documentclass[
reprint,
superscriptaddress,
aps,
pre,
]{revtex4-1}

\usepackage{graphicx}
\usepackage{dcolumn}
\usepackage{bm}
\usepackage[colorlinks=true,linkcolor=Blue,citecolor=Blue,urlcolor=Blue]{hyperref}
\usepackage[utf8]{inputenc}
\usepackage{amsmath, amsthm, amssymb, amsfonts}
\usepackage{accents}
\usepackage[usenames,dvipsnames]{xcolor}
\usepackage[caption=false]{subfig}
\usepackage{tikz}
\usepackage{array}
\usepackage{dsfont}
\usepackage{mathtools}
\usepackage{soul,xcolor}
\usepackage[capitalise]{cleveref}

\begin{document}

\newcommand{\mytitle}{Critical Switching in Globally Attractive Chimeras}
\title{\mytitle}

\author{Yuanzhao Zhang}
\affiliation{Department of Physics and Astronomy, Northwestern University, Evanston, Illinois 60208, USA}
\author{Zachary G. Nicolaou}
\affiliation{Department of Physics and Astronomy, Northwestern University, Evanston, Illinois 60208, USA}
\author{Joseph D. Hart}
\affiliation{Institute for Research in Electronics and Applied Physics, University of Maryland, College Park, Maryland 20742, USA}
\affiliation{Department of Physics, University of Maryland, College Park, Maryland 20742, USA}
\author{Rajarshi Roy}
\affiliation{Institute for Research in Electronics and Applied Physics, University of Maryland, College Park, Maryland 20742, USA}
\affiliation{Department of Physics, University of Maryland, College Park, Maryland 20742, USA}
\affiliation{Institute for Physical Science and Technology, University of Maryland, College Park, Maryland 20742, USA}
\author{Adilson E. Motter}
\affiliation{Department of Physics and Astronomy, Northwestern University, Evanston, Illinois 60208, USA}
\affiliation{Northwestern Institute on Complex Systems, Northwestern University, Evanston, Illinois 60208, USA}

\begin{abstract}
We report on a new type of chimera state that attracts almost all initial conditions and exhibits power-law switching behavior in networks of coupled oscillators. 
Such {\it switching chimeras} consist of two symmetric configurations, which we refer to as subchimeras, in which one cluster is synchronized and the other is incoherent.
Despite each subchimera being linearly stable, switching chimeras are extremely sensitive to noise: Arbitrarily small noise triggers and sustains persistent switching between the two symmetric subchimeras.
The average switching frequency scales as a power law with the noise intensity, which is in contrast with the exponential scaling observed in typical stochastic transitions.
Rigorous numerical analysis reveals that the power-law switching behavior originates from intermingled basins of attraction associated with the two subchimeras, which, in turn, are induced by chaos and symmetry in the system. 
The theoretical results are supported by experiments on coupled optoelectronic oscillators, which demonstrate the generality and robustness of switching chimeras.
\vspace{3mm}

\noindent DOI: \href{https://doi.org/10.1103/PhysRevX.10.011044}{10.1103/PhysRevX.10.011044} 
\end{abstract}


\maketitle

\section{Introduction}

The relationship between symmetry and synchronization underlies many recent discoveries in network dynamics. 
Symmetries influence the possible dynamical patterns in a network \cite{pecora2014cluster,golubitsky2016rigid} and can either facilitate \cite{nicosia2013remote,okuda1991mutual,zhang2017incoherence} or inhibit  \cite{PhysRevLett.117.114101,saa2018symmetries,hart2019topological} synchronization.
A particularly interesting symmetry phenomenon in networks is the coexistence of coherent and incoherent clusters in populations of identically coupled identical oscillators \cite{kuramoto2002coexistence,kaneko1990clustering}---the so-called chimera states \cite{abrams2004chimera}.
Since chimeras have less symmetry than the system itself, they represent symmetry-broken states \cite{crawford1991symmetry} of the network dynamics.
Over the years, different forms of chimera states have been discovered \cite{sethia2008clustered,abrams2008solvable,martens2010solvable,larger2013virtual,yeldesbay2014chimeralike,zakharova2014chimera,xie2014multicluster,semenova2016coherence,shena2017turbulent}, which has been accompanied by new results on robustness \cite{hagerstrom2012experimental,tinsley2012chimera,martens2013chimera,bick2017robust,totz2018spiral} and existence conditions \cite{omel2008chimera,sethia2014chimera,ashwin2015weak,martens2016basins,nicolaou2017chimera,omel2018mathematics,bansal2019cognitive}.

Early work on chimera states focused mainly on networks of phase oscillators in the limit of a large system size \cite{panaggio2015chimera}, where dimension reduction is often possible by employing the Ott-Antonsen ansatz \cite{ott2008low,ott2009long,pazo2014low}.
For finite-size systems, some chimera states have been shown to be long transients \cite{wolfrum2011chimera}, while others have been shown to be stable \cite{pikovsky2008partially,panaggio2016chimera} using the Watanabe-Strogatz ansatz \cite{watanabe1994constants,marvel2009identical}. 
Recent research has placed an increased emphasis on chimeras in finite-size networks of chaotic oscillators \cite{omelchenko2011loss,omelchenko2012transition,semenova2015does,hart2016experimental,cho2017stable}, which are important given the prevalence of chaos in physical systems \cite{boccaletti2002synchronization}.
In that context, it has been shown that the stability of chimera states can be studied rigorously using cluster synchronization techniques \cite{hart2016experimental,cho2017stable}.

Even for permanently stable chimeras, an important question is how carefully one has to prepare the initial conditions in order to observe them. 
Early examples of chimera states required specially prepared initial conditions \cite{abrams2004chimera,abrams2008solvable,martens2010bistable}, while more recent examples include chimera states that emerge from a wide range of initial conditions \cite{omel2008chimera,sethia2013amplitude,schmidt2014coexistence,yeldesbay2014chimeralike,schmidt2015clustering,kotwal2017connecting}. 
In the presence of global feedback control, some chimeras have even been observed to attract almost all initial conditions \cite{bordyugov2010self,sieber2014controlling}. 
However, whether globally attractive chimeras can emerge in the absence of control is still an open problem.

Because of the symmetry-broken nature of chimera states, another important question concerns the coexistence of multiple chimeras \cite{martens2010bistable} and the possibility of transitions between them \cite{ma2010robust}. 
When multiple chimeras coexist, adding fluctuation or mismatch terms may induce switching events between them. 
This phenomenon has been studied under the name of ``alternating chimeras'' \cite{laing2012disorder,buscarino2015chimera,semenova2016coherence}. 
In previous studies, finite transition barriers must be overcome for transitions between otherwise persistent chimeras to occur. 
Accordingly, the transition rates are expected to scale exponentially with noise intensity.

\begin{figure*}[htb!]
\centering
\includegraphics[width=1\linewidth]{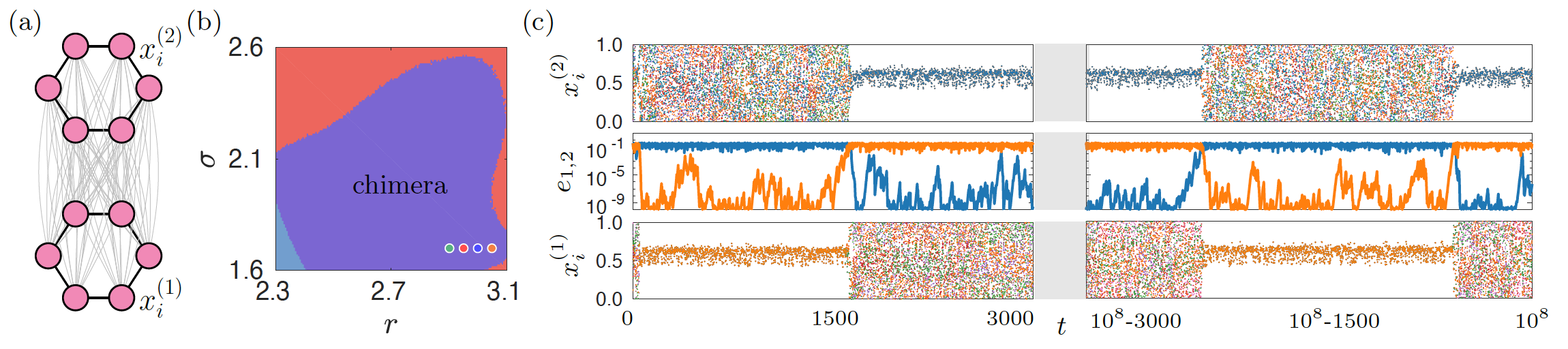}
\vspace{-5mm}
\caption{
Globally attractive chimera state whose coherent and incoherent clusters switch under extremely small noise.
(a) Network system, formed by two rings of logistic maps mutually coupled through weaker links [\cref{eq:0}]. 
(b) Parameter space color coded according to the linear stability of the possible states, namely, whether both rings can synchronize (cyan), only one ring can synchronize (purple), or neither ring can synchronize (red). The four dots mark the parameters used in \cref{fig:2}.
(c) Direct simulation of the system for $\sigma=1.7$ and $r=3.05$ [orange dot in (b)] for noise intensity $\xi = 10^{-10}$, illustrating the dynamics of a switching chimera. The top and bottom panels show the oscillator states in each of the two rings (color coded by oscillator, where single-color segments indicate synchronization), while the center panel shows the synchronization error [defined in \cref{eq:error}] in each ring.
}
\label{fig:1}
\end{figure*}

Here, we report on switching chimeras, which are chimera states that both exhibit power-law dependence of the switching frequency on noise intensity and attract almost all initial conditions in the absence of control.
A switching chimera is comprised of two symmetric metastable states---referred to as subchimeras---between which the switching occurs.  
The power-law switching dynamics is a signature of critical behavior and stems from a vanishing quasipotential barrier between the two metastable states. 
It follows that the switching persists indefinitely for any nonzero noise intensity.
Strikingly, when the noise intensity is strictly zero, the symmetric subchimeras are linearly stable. 
Thus, the deterministic dynamics settle into one of the two subchimeras, and, as in the original studies of chimeras, the state symmetry is broken. 
For any nonzero noise intensity, however, the long-term dynamical symmetry is restored due to the persistent switching between the two subchimeras.
This dependence on noise intensity shares similarities with singular limits \cite{berry2002singular}, in that the asymptotic network dynamics are qualitatively different for zero and small noise.
Our analytical and numerical results are further validated by an experimental demonstration of switching chimeras in networks of optoelectronic oscillators.
We suggest that switching chimeras can find applications in the study of intermittently alternating dynamics in biological systems and the development of approaches to measure small experimental noise.

The paper is organized as follows.
In Sec.~\ref{sec:appetizer}, we introduce a representative system exhibiting switching chimeras.
The power-law dependence between the average switching period and noise intensity is presented in Sec.~\ref{sec:power-law}.
This critical switching behavior is then established and explained from various angles in the subsequent subsections. 
In Sec.~\ref{sec:log space}, we show that it arises robustly in a first-exit model derived from an extension of the Freidlin-Wentzell theory.
In Sec.~\ref{sec:pathways}, we further elucidate the mechanism underlying the switching dynamics by describing the dominant transition paths and the role of invariant saddles.
In Sec.~\ref{sec:quasi-potential}, we relate the scaling in the switching dynamics with the existence of transition paths of arbitrarily small action and compare it to critical phenomena in phase transitions.
In Sec.~\ref{sec:riddled}, we establish a connection between power-law switching and intermingled basins of attraction.
Experiments confirming switching chimeras and their power-law scaling in a network of optoelectronic oscillators are presented in Sec.~\ref{sec:experiment}.
In Sec.~\ref{sec:connections}, we discuss connections between switching chimeras and other phenomena in physical and biological systems.
Finally, we present our concluding remarks in Sec.~\ref{sec:discussions}.

\section{Computational observation of switching chimeras}
\label{sec:appetizer}

We consider $2n$-node networks formed by two rings of $n$ nodes, with nearest-neighbor coupling of strength $\sigma$ in each ring. 
The two rings are all-to-all coupled by weaker links of strength $c\sigma$ for some $0<c<1$.
In this way, all the nodes are identically coupled, as shown by the network diagram in \cref{fig:1}(a).
We assume the oscillators are diffusively coupled, so the network can be represented through a Laplacian matrix in the dynamical equation.
Adding to each node an uncorrelated Gaussian noise  term of zero mean and tunable standard deviation $\xi$  (which we refer to as the noise intensity) and writing down the coupling explicitly, the resulting stochastic dynamical equation for the first ring reads:
\begin{equation}
  \begin{split}
    x_i^{(1)}[t+1] = & \Big\{ \underbrace{\,r\,f\big(x_i^{(1)}[t]\big)}_\text{intrinsic dynamics} \\
    + & \underbrace{\sigma \left( f\big(x_{i-1}^{(1)}[t]\big) + f\big(x_{i+1}^{(1)}[t]\big) - 2f\big(x_i^{(1)}[t]\big) \right)}_\text{intracluster coupling} \\ 
    + & \underbrace{c\sigma \sum_{j=1}^n \left( f\big(x_{j}^{(2)}[t]\big) - f\big(x_i^{(1)}[t]\big) \right)}_\text{intercluster coupling} \\
    + & \underbrace{\xi N_i^{(1)}[t]}_\text{Gaussian noise} \Big\} \,\, \text{mod} \, 1, \qquad 1\leq i \leq n,
\end{split}
\label{eq:0}
\end{equation}
where $N_i^{(1)}$ is Gaussian noise with unit standard deviation and the superscripts indicate which ring the variables are associated with. 
The dynamical equation for the second ring can be expressed similarly. 
(We note that it is not essential for the dynamics to be discrete; an example of switching chimeras in systems with continuous-time dynamics is presented in Supplemental Material \cite{SM}, Sec.~S5.)

We first assume that the dynamics of each node is governed by a logistic map $f(x) = x(1-x)$.
For concreteness, we also set $n=6$ and $c=0.2$ unless mentioned otherwise.
Using a generalization of the master stability function formalism developed in Ref.~\cite{hart2019topological}, we can calculate the maximum transverse Lyapunov exponent associated with chimera states efficiently (\cref{sec:stability}).
In particular, we find  parameters under which
\begin{enumerate}
	\item[i)] the two clusters cannot be simultaneously in stable synchronous states (i.e., any solution satisfying $x_i^{(1)}[t]=s_1[t]$, $x_i^{(2)}[t]=s_2[t]$ for all $i$ is linearly unstable);
	\item[ii)] one of the clusters can be in a stable synchronous state if the other cluster is not.
\end{enumerate}
Inside the region where both conditions are satisfied, {\it coherence is induced by incoherence}, meaning that synchronization in one cluster is stabilized by desynchronization in the other cluster. \Cref{fig:1}(b) shows that the system in \cref{fig:1}(a) has a large parameter region (purple) in which these two conditions are satisfied. In that region, chimera states are linearly stable and do not coexist with stable globally synchronized states.

To confirm that the desynchronized ring is indeed in an incoherent state, we run direct simulations \footnote{Simulation code available at \url{https://github.com/y-z-zhang/switching-chimeras}} from random initial conditions for $10^8$ iterations under noise of intensity $\xi=10^{-10}$.
\Cref{fig:1}(c) shows representative trajectories and associated synchronization errors for $\sigma = 1.7$ and $r = 3.05$.
The synchronization error in the $j$-th cluster is defined as 
\begin{equation}
  e_j \coloneqq \sqrt{\sum_{i=1}^n \frac{\|x_i^{(j)} - \bar{x}^{(j)}\|^2}{n}},
  \label{eq:error}
\end{equation} 
where $\| x \| = \min(|x|,1-|x|)$ and $\bar{x}^{(j)}$ is the mean of $x_i^{(j)}$ over all $i$.

The system exhibits not only chimera dynamics but also persistent transitions in which the coherent and incoherent rings switch roles: As one ring loses synchrony and becomes incoherent, the other ring synchronizes.
Moreover, as we show below, the switching observed here is critical---the transition rate depends on the noise intensity as a power law and switching can be triggered by arbitrarily small noise.
This power-law dependence distinguishes switching chimeras from previously reported ``alternating chimeras,'' in which the transitions either are forced by large fluctuation terms \cite{ma2010robust,laing2012disorder,buscarino2015chimera,semenova2016coherence} or rely on heteroclinic dynamics \cite{haugland2015self,bick2018heteroclinic,goldschmidt2019blinking}.
In the first case, there are finite barriers separating the different states, while in the second case each state is inherently unstable and switching occurs in the absence of noise. 

The persistence of switching chimeras under many transition cycles suggests it is globally attractive. 
To verify that this is indeed the case, we evolve the system for $10^4$ iterations starting from $10^6$ different random initial conditions for $\sigma = 1.7$ and $r = 2.9,\,2.95,\,3.0,\text{ and }\,3.05$ [dots in \cref{fig:1}(b)]. 
In all tests, the oscillators are swiftly attracted to the chimera state and no other attractors are observed. 
Further evidence of this global attractiveness is presented in Supplemental Material \cite{SM}, Sec.~S1, where we also demonstrate the prevalence of switching chimeras across different cluster sizes, intercluster coupling strengths, and intracluster coupling range.

\section{Power-law switching}

\subsection{Extreme sensitivity to noise}
\label{sec:power-law}

\begin{figure}[t]
\centering
\includegraphics[width=.85\columnwidth]{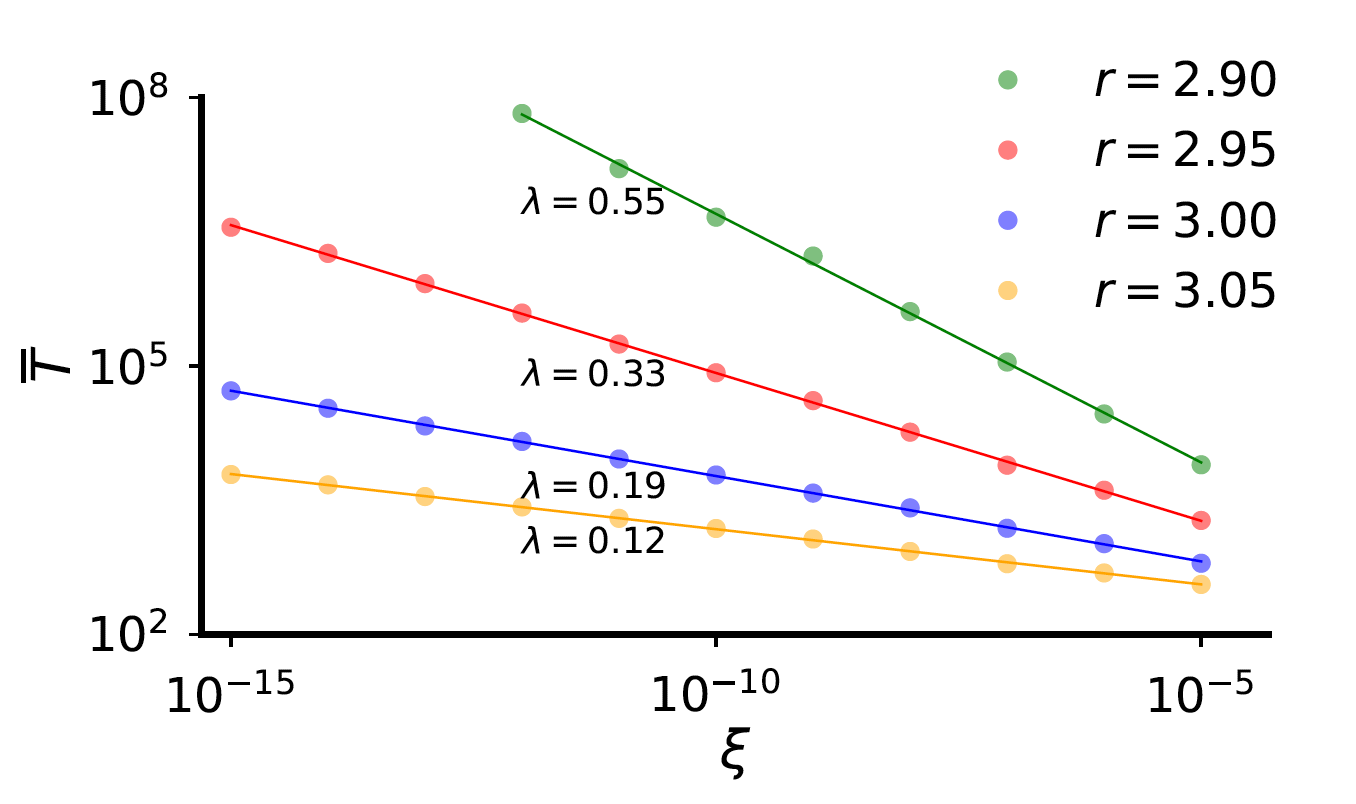}
\vspace{-3mm}
\caption{Average switching period $\overline{T}$ as a function of the noise intensity $\xi$ for $\sigma=1.7$ and various values of $r$ [dots in \cref{fig:1}(b)]. The switching periods are extracted from long time series of switching chimeras obtained by simulating \cref{eq:0} for different values of $\xi$. The numbers indicate the scaling exponents and are obtained through least-square fit (slopes of the solid lines).}
\label{fig:2}
\end{figure}

Next, we present numerical results characterizing the effect of noise intensity $\xi$ on the average switching period $\overline{T}$.
\Cref{fig:2} shows that, as one approaches the boundary of the chimera region [from the green dot to the orange dot in the bottom right of \cref{fig:1}(b)], $\overline{T}$ decreases and switching becomes more frequent.
For each fixed value of $r$, the average switching period increases as the noise intensity decreases, with scaling that follows a power law.
It is remarkable that even noise of intensity as small as $\xi=10^{-15}$ (the resolution limit of computers using double-precision floating-point format) can induce frequent switching.

This switching between the coherent and incoherent clusters does not contradict the fact that synchronization in one cluster is linearly stable if the other cluster is incoherent. 
This is the case because linear stability analysis assumes the perturbations to be infinitesimally small, whereas finite-size perturbations, no matter how small, can still grow large enough along the unstable portions of a chaotic attractor to disrupt synchrony in the coherent ring and induce switching.

\begin{figure*}[!hbt]
\centering
\includegraphics[width=1\linewidth]{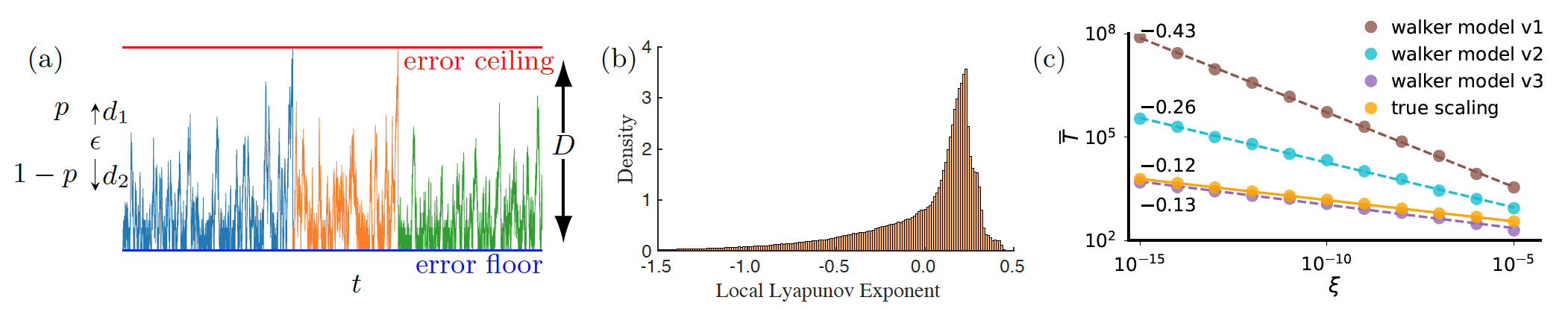}
\vspace{-5mm}
\caption{Modeling transitions in switching chimeras.
(a) Illustration of a random walk model in the log-error space, where a switching event is triggered when the walker reaches the error ceiling. The time series is colored differently after each switching event.
(b) Distribution of the local Lyapunov exponents associated with \cref{eq:0} for $\sigma = 1.7$ and $r = 3.05$, which is used to refine the random walk model for the switching chimeras.
(c) Power-law scalings predicted by the random walk model and its refined versions (dashed lines). The scaling obtained from direct simulations of \cref{eq:0} is also shown for comparison (solid orange line).}
\label{fig:3}
\end{figure*}

The power-law scaling of the average switching period and, consequently, the extreme noise sensitivity of chimera states, makes the switching behavior observed here ``anomalous'' in the sense that it appears to contradict the Freidlin-Wentzell theory \cite{freidlin1998random}.
According to that theory, for a stochastic system with deterministic dynamics $\bm{F}$ and a noise term of intensity $\xi$,
\begin{equation}
  \bm{x}[t+1] = \bm{F}(\bm{x}[t]) + \xi \bm{N}[t],
\end{equation}
the rate of transition from one metastable \footnote{We consider a state to be metastable if it is linearly stable in the absence of noise but only has a finite lifetime when noise is present.} state $\mathcal{A}$ to another metastable state $\mathcal{B}$ scales as $\exp(-S_{\mathcal{A}\rightarrow\mathcal{B}}/\xi^2)$, and the first exit time scales as $\exp(S_{\mathcal{A}\rightarrow\mathcal{B}}/\xi^2)$ \cite{wells2015control}. 
Here, $S_{\mathcal{A}\rightarrow\mathcal{B}}$ is the infimum of the Freidlin-Wentzell action among all paths $\bm{X}$ connecting state $\mathcal{A}$ to state $\mathcal{B}$:
\begin{equation}
	S_{\mathcal{A}\rightarrow\mathcal{B}} \coloneqq \frac{1}{2} \inf_{\substack{\bm{X} \\ \bm{X}[0]\in\mathcal{A} \\ \bm{X}[m]\in\mathcal{B}}} \sum_{t=0}^{m-1} \|\bm{X}[t+1] - \bm{F}(\bm{X}[t])\|^2.
\end{equation}
The infimum of the action measures how much one has to work against the deterministic part of the dynamics to induce a transition from $\mathcal{A}$ to $\mathcal{B}$. 
This quantity is also known in the literature as a quasipotential barrier \cite{zhou2012quasi} and is analogous to a potential barrier for transitions in gradient systems.

\subsection{First-exit problem in log-error space}
\label{sec:log space}

Although the power-law scaling observed for switching chimeras and the exponential scaling predicted by the Freidlin-Wentzell theory seem incompatible at first glance, we can establish a connection between them.
We first note that the synchronization error inside the coherent ring usually fluctuates close to an error floor determined by the noise intensity, but switching can be triggered by rare events that drive the error all the way to an error ceiling determined by the synchronization error of the incoherent ring [for an example, see the middle panel of \cref{fig:1}(c)].
Moreover, since the variational equation acts multiplicatively on the synchronization error see \cref{sec:stability}), the error naturally evolves on a log scale as long as the linearization around the synchronization manifold is still valid.

Motivated by these observations, we focus on an attribute $\epsilon$, defined as the logarithm of the synchronization error inside the coherent ring:
\begin{equation}
  \epsilon \coloneqq \ln\left(\min\{e_1,e_2\}\right).
\end{equation}
As a first approximation, the dynamics of $\epsilon$ can be modeled as a biased one-dimensional random walk confined within two boundaries, corresponding to the error floor and the error ceiling.
At each step, $\epsilon$ has probability $p$ of moving up a fixed distance $d_1$ and probability $1-p$ of moving down a distance $d_2$.
The random walker starts from the error floor, and it never goes below that boundary. 
Every time $\epsilon$ reaches the error ceiling, we consider that a switching event has occurred and reset $\epsilon$ to the lower boundary.
An illustration of this process can be found in \cref{fig:3}(a).

To derive a relation between the average switching period $\overline{T}$ and the interboundary distance $D$ in the random walk model, we note that when $pd_1<(1-p)d_2$ and $D \gg d_{1,2}$ this is a first-exit problem. 
Thus, according to the Freidlin-Wentzell theory,
\begin{equation}
  \overline{T} \propto \exp(\lambda D),
  \label{eq:1}
\end{equation}
where $\lambda$ is some constant determined by $p$, $d_1$, and $d_2$.
Now recall that $D$ is determined by the distance between the error floor and error ceiling.
The error floor is given by $\ln(\xi)$, and, without loss of generality, we set the error ceiling to be $1$.
Thus, $D = \ln(1) - \ln(\xi) = \ln(\xi^{-1})$, and \cref{eq:1} becomes
\begin{equation}
  \overline{T} \propto \xi^{-\lambda}.
  \label{eq:2}
\end{equation}
This scaling reproduces the power-law relation between the average switching period $\overline{T}$ and the noise intensity $\xi$ observed in \cref{fig:2}.

\begin{figure*}[!hbt]
\centering
\includegraphics[width=.9\linewidth]{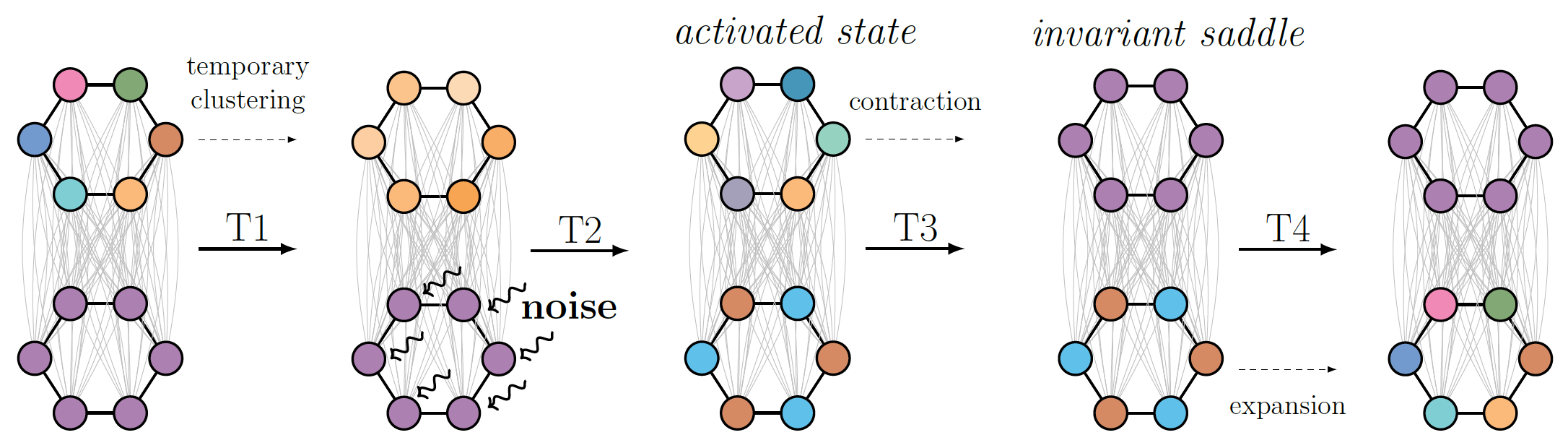}
\caption{
Dominant transition pathway between the two symmetric subchimeras, which consists of the intermediary stages T1 to T4. Only T2 requires activation from noise, which can be arbitrarily small but not strictly zero; all other transitions follow directly from the deterministic dynamics of \cref{eq:0}.
In particular, T3 and T4 follow the stable and unstable manifolds of the invariant saddle, respectively.
}
\label{fig:4}
\end{figure*}

We now turn to a more quantitative analysis to support the idea that the switching events in the original system can be inferred from the one-dimensional attribute $\epsilon$.
Starting with the system in \cref{eq:0}, we compute the growth rate of the synchronization error in the coherent ring $\epsilon[t+1] - \epsilon[t]$ at each iteration.
The distribution of this quantity, which we call the {\it local} Lyapunov exponent, is shown in \cref{fig:3}(b) for $\sigma = 1.7$ and $r = 3.05$.
Of all the local Lyapunov exponents sampled, 35\% are negative, with a mean of $-0.46$; the remaining $65\%$ of the exponents are positive, with a mean of $0.19$.
Because $e$ is a one-dimensional variable, the Lyapunov exponent that determines its asymptotic stability at $0$ is given by averaging over the local Lyapunov exponents from $t=0$ to $t=\infty$.
Since $-0.46 \times 0.35 + 0.19 \times 0.65 < 0$, although $65\%$ of the chaotic attractor is repelling, the chimera state is actually linearly stable.
From the above information, we can set $p = 0.65$, $d_1 = 0.19$, and $d_2 = 0.46$ in our random walk model and calculate the relation between the average switching period $\overline{T}$ and the noise intensity $\xi$.

The brown circles in \cref{fig:3}(c) indicate how $\overline{T}$ scales with $\xi$ for this random walk model; they follow a well-defined power law, as expected from \cref{eq:2}. 
But it is also clear that a random walk is not a very accurate picture for the dynamics of $\epsilon$, since the predicted average switching periods are much larger than the ones obtained from simulating \cref{eq:0} (orange circles).
This discrepancy is partially due to the crude approximation we made when fixing the step sizes of the random walk to be constants.
If we choose the step size as well as the direction of the random walk according to the distribution in \cref{fig:3}(b), we observe the scaling indicated by cyan circles in \cref{fig:3}(c), which is closer to the true scaling.
However, the predicted exponent of $-0.26$ is not yet close to the true value of $-0.12$, which indicates that something is still missing.

The approach we just took is equivalent to shuffling the time series of the local Lyapunov exponents and using the shuffled sequence to generate the random walk.
This shuffling preserves the information of the full distribution but ignores temporal correlations.
Because the stable and unstable portions of a chaotic attractor are usually not well mixed, the actual evolution of $\epsilon$ is a non-Markovian process, and we expect the temporal information to be relevant.
This effect tends to correlate the upward movements of $\epsilon$, which, in turn, makes it more likely for $\epsilon$ to reach the error ceiling and shortens the average switching period for small noise.
When the temporal information is incorporated into the model (by using the original sequence of local Lyapunov exponents rather than randomly sampling them), we arrive at a more realistic model for the switching dynamics, which takes the form of a deterministic walker.
The prediction of this refined model (purple circles) is in excellent agreement with the true scaling (orange circles).

It is important to note that the power-law scaling is preserved even after we allow variable step sizes and strong correlation between steps in our model.
We thus suggest that \cref{eq:2} is robust and that power-law switching is expected for a general class of systems. 
Transitions in such systems can be modeled as a first-exit problem in which the distance to the exit increases linearly with the logarithm of the inverse of noise intensity.

\subsection{Transition pathways}
\label{sec:pathways}

\begin{figure}[!bt]
\centering
\includegraphics[width=1\columnwidth]{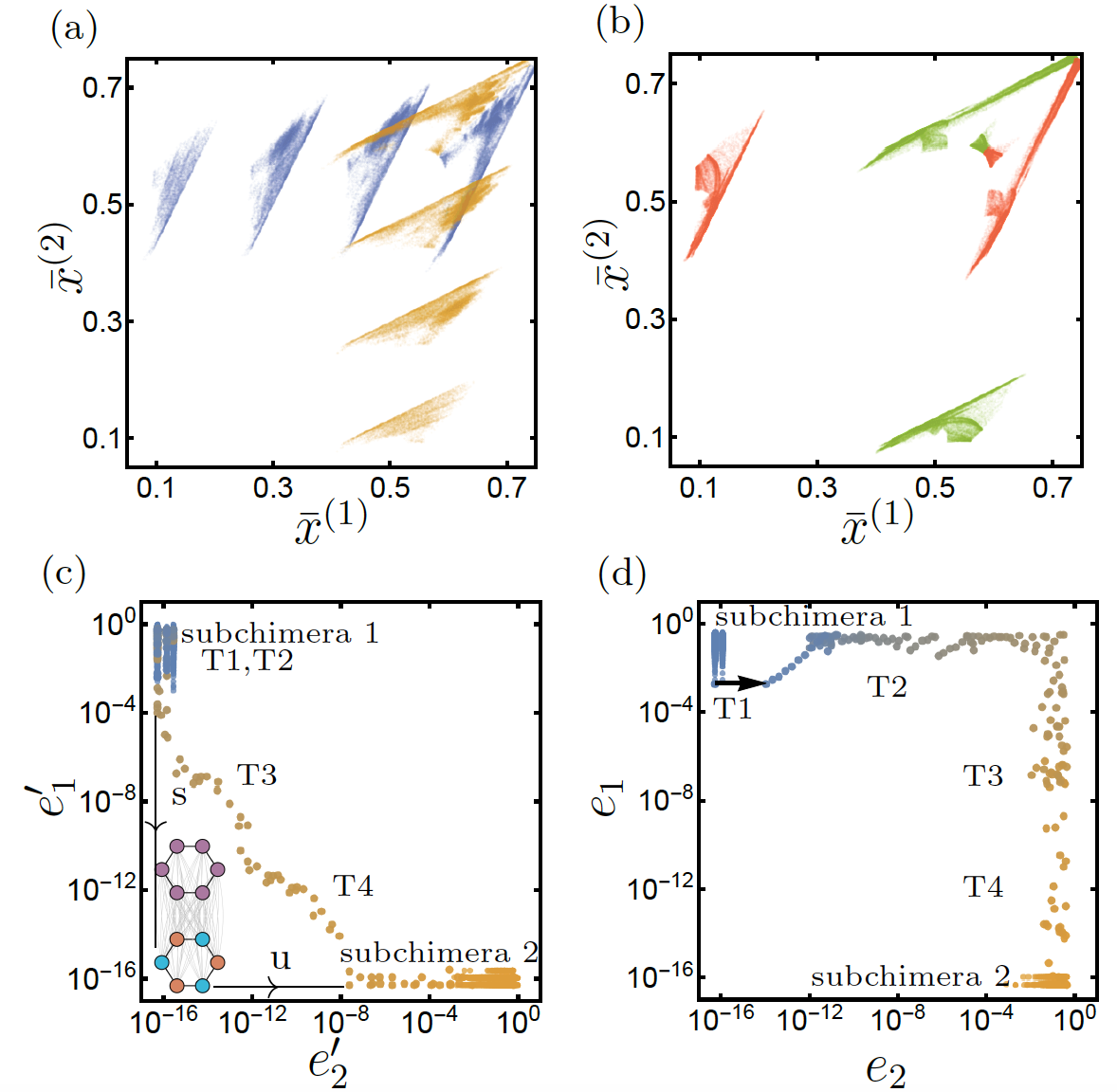}
\vspace{-5mm}
\caption{Projections of invariant sets and transition paths.
(a) Symmetric subchimeras when projected onto the mean state of each ring. Each subchimera is indicated by a different color.
(b) Invariant saddle in \cref{fig:4} (and its symmetric counterpart) projected onto the mean state of each ring.  
(c) Transition path with an action of $10^{-28}$ projected onto coordinates $e'_1$ and $e'_2$.  Under this projection, the invariant saddle is projected onto the lower left corner. The stable and unstable manifolds of the invariant saddle are marked by s and u, respectively. The path starts at the blue subchimera in the upper left corner and ends at the orange subchimera in the lower right corner.
(d) Same transition path as in (c) projected onto $e_1$ and $e_2$. The perturbation that initiates the transition is marked by an arrow.}
\label{fig:5}
\end{figure}

We can gain a deeper understanding of the switching dynamics by investigating the transition paths connecting the two symmetric subchimeras. 
One natural question concerns whether there is a single pathway or multiple pathways for the switching. 
If multiple pathways exist, do they intersect at key intermediate states?
For the system in \cref{fig:1}(a), with $n=6$, it turns out that there is only one dominant pathway when noise is small. 
We illustrate the key transitions (T1 to T4) and intermediate states of this pathway in \cref{fig:4}. We later analyze an explicit realization of this pathway in \cref{fig:5}, which provides strong numerical support for the following transition sequence: \\
(T1) Starting from one of the subchimeras, the incoherent ring occasionally visits near-synchronized states (referred to as temporary clustering in \cref{fig:4}). \\
(T2) The temporary clustering in the incoherent ring strongly correlates with the instability windows in the coherent ring.
This correlation is not surprising, since states with both rings synchronized are unstable.
Within those short windows, small noise or perturbations applied to the coherent ring are amplified and lead to a short-wavelength bifurcation.
That is, the coherent ring partially desynchronizes and splits into two alternating groups with different dynamics (oscillators in the same group remain synchronized).
Reaching this ``activated state'' is the only stage in which noise is needed, even though it can be arbitrarily small.\\
(T3, T4) The state between T3 and T4 lives in an invariant subspace induced by the rotational symmetry in each ring. 
In fact, the state is an invariant saddle and serves as the key intermediate state connecting the two subchimeras. 
During T3, the system moves along the stable manifold of the invariant saddle, and the six oscillators in the upper ring converge to a synchronized state.
During T4, the system moves away from the saddle following its unstable manifold, where the partially desynchronized state in the lower ring evolves into an incoherent state. 
The roles of the rings are now reversed, thus concluding the entire sequence of transitions from one subchimera to the other.

The short-wavelength perturbation
\begin{equation}
  \bm{\Delta}_{sw}(\delta) = \frac{1}{\sqrt{6}}(\delta,-\delta,\delta,-\delta,\delta,-\delta),
\end{equation}
where the $i$-th component of this vector is to be interpreted as a perturbation to the $i$-th node in the ring, is the dominant instability in the coherent ring according to our linear stability analysis and is the one being excited by noise during transition T2. 
To further support this claim, we run direct simulations of \cref{eq:0}, but with $\bm{\Delta}_{sw}$ filtered out from the noise applied to each ring. 
This time, for noise intensity $\xi \leq 10^{-9}$, the average switching period $\overline{T}$ becomes independent of $\xi$ and always equals the average switching period induced by round-off errors, as shown in \cref{sec:swb route}.
These simulations confirm that the overwhelming majority of the switching events must be initiated through a short-wavelength bifurcation in the coherent ring when noise is small \footnote{The same result holds for all $n>2$. A ring network with $n$ nodes has eigenvalues $\lambda_k = 4\sin^2(k\pi/n)$ and eigenvectors $\bm{\eta}_k = (1,e^{\frac{2\pi\mathrm{i}}{n}k},e^{\frac{2\pi\mathrm{i}}{n}2k},\dots,e^{\frac{2\pi\mathrm{i}}{n}(n-1)k})/\sqrt{n}$. For \cref{eq:0}, the leading instability is associated with the largest eigenvalue. This corresponds to $\bm{\eta}_{n/2} = (1,-1,1,-1,\dots,1,-1)/\sqrt{n}$ for $n$ even and to both $\bm{\eta}_{(n-1)/2}$ and $\bm{\eta}_{(n+1)/2}$ for $n$ odd.}.

To better visualize the subchimeras and the invariant saddles, we project them onto the mean state of each ring: $\bar{x}^{(1)} = \sum x_i^{(1)}/n$ and $\bar{x}^{(2)} = \sum x_i^{(2)}/n$.
\Cref{fig:5}(a) shows the projection of the two symmetric subchimeras colored in blue and orange, respectively.
We can see the fine structure of the subchimeras under this projection, which is indicative of their fractal nature.
In \cref{fig:5}(b), we show the projection of the two invariant saddles (red and green).

We now try to explicitly find a least-action path connecting the two subchimeras, which can be challenging even for transitions between fixed points or periodic orbits \cite{wells2015control,zhou2012quasi}.
In our case, the high dimensionality and the chaotic nature of the subchimeras make the optimization of the transition path extremely difficult when using traditional methods.
Fortunately, the mechanism presented in \cref{fig:4} points to an efficient way of finding paths of arbitrarily low action connecting the two subchimeras.
We simply wait for the incoherent ring to visit a near-synchronized state and then introduce a one-time perturbation in the form of $\bm{\Delta}_{sw}(\delta)$ to excite the short-wavelength bifurcation in the coherent ring.
If a transition is successfully triggered, the action of the transition path is simply $\frac{1}{2}\delta^2$.

Using this strategy, we can easily find a transition path with action as small as $10^{-28}$ (i.e., $\delta$ around $10^{-14}$), which is shown in \cref{fig:5}(c) and \cref{fig:5}(d) for different projections.
The coordinate $e'_1$ ($e'_2$) in \cref{fig:5}(c) is defined as the sum of the synchronization error among the odd oscillators and the synchronization error among the even oscillators in the first (second) ring.
For this projection, the two subchimeras are found in the upper left and the lower right corners, while the key invariant saddle connecting the two subchimeras is projected onto the lower left corner ($e'_1=e'_2=0$).
It is informative to view the projected transition path  in \cref{fig:5}(c) in light of the pathway shown in \cref{fig:4}: The first two transitions (T1 and T2) correspond to the upper left corner, while the other two transitions (T3 and T4) loop around the lower left corner as they follow the stable and unstable manifolds of the invariant saddle closely.
Conversely, the projected path provides strong numerical support for the pathway illustrated in \cref{fig:4}.
However, the evidence is not yet conclusive, as states with both rings synchronized also project onto the lower left corner for the coordinates in \cref{fig:5}(c).
Could the two subchimeras be connected by an unstable synchronized state instead of the invariant saddles in \cref{fig:4}?
The projection to the synchronization errors $e_1$ and $e_2$ in \cref{fig:5}(d) excludes this possibility, since the path goes through the upper right corner (both rings desynchronized) rather than the lower left corner (each ring synchronized).
Multiple transition paths with action ranging from $10^{-30}$ to $10^{-10}$ are tested, and they are all qualitatively identical to each other under both projections.
This evidence further supports the existence of a dominant transition pathway for the observed switching between subchimeras.

\subsection{Connections with critical phenomena}
\label{sec:quasi-potential}

\begin{figure}[t]
\centering
\includegraphics[width=1\columnwidth]{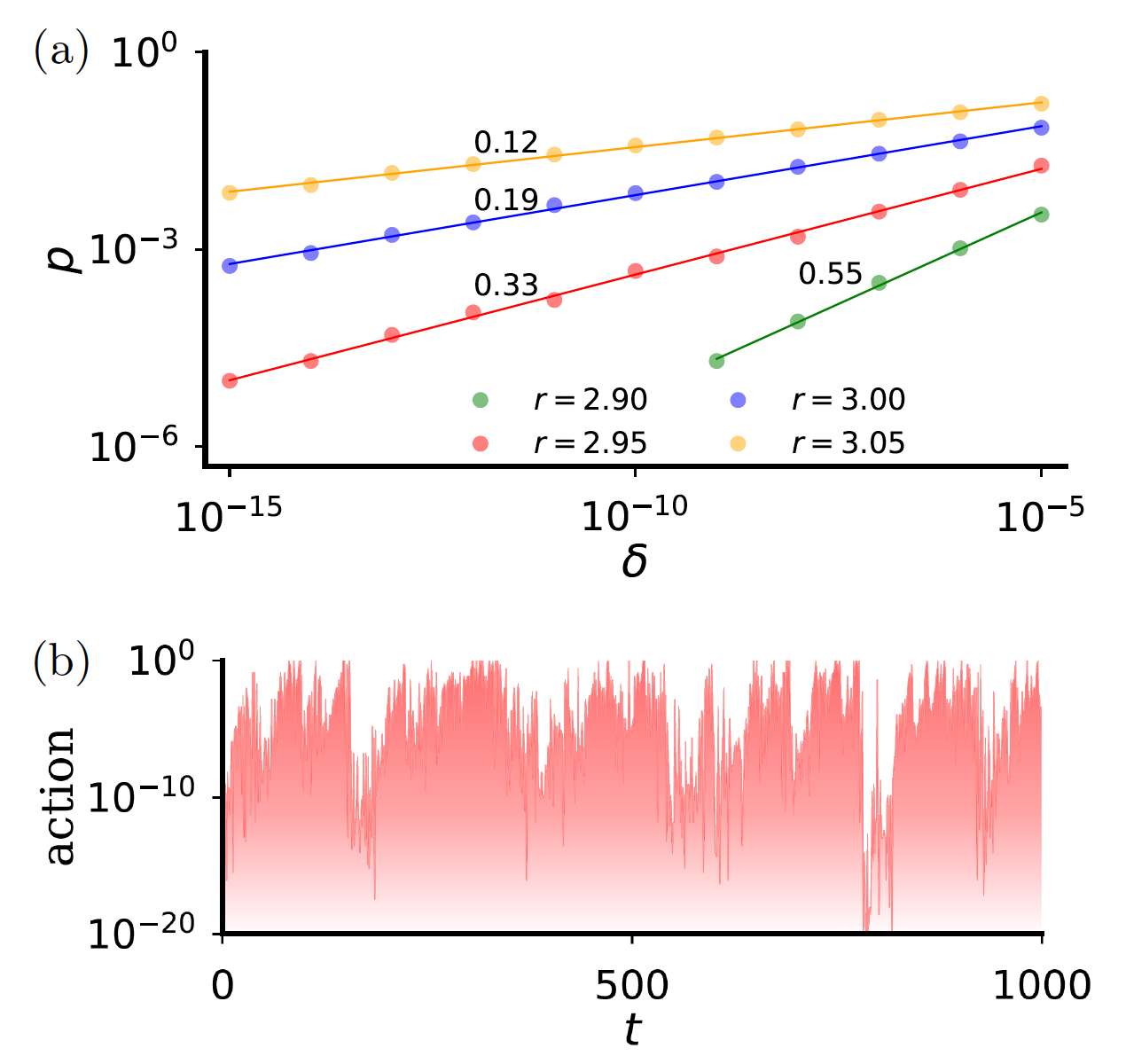}
\vspace{-5mm}
\caption{
Action profile for transition paths.
(a) Probability $p$ of finding small-action transition paths by introducing a short-wavelength perturbation of magnitude $\delta$ in a single iteration.
The simulations are performed for $\xi = 0$ and the other parameters are the same as in \cref{fig:2}.
Paths with arbitrarily small action exist but small-action paths become increasingly more difficult to find as the available action is decreased, 
resulting in power-law relationships between the probability $p$ and the perturbation size $\delta$.
Notice that the scaling exponents here match those in \cref{fig:2}.
(b) Minimum action ($\frac{1}{2}\delta^2$) needed to induce a transition by applying $\bm{\Delta}_{sw}(\delta)$ at a given time $t$, for $\xi = 0$, $\sigma = 1.7$, and $r = 2.95$. 
This highly structured profile can be regarded as a visualization of the transition-barrier landscape for switching chimeras.
}
\label{fig:6}
\end{figure}

The fact that switching can be induced by arbitrarily small noise but not in the absence of noise implies that (i) no matter how small the action of a transition path, we can always find another path with even smaller action, and (ii) there is no zero-action path of finite length connecting the two subchimeras.
Thus, a least-action path does not exist in our system. 
Instead, given an arbitrarily small upper bound on the available action, there are always finite-length transition paths that meet that constraint. 
It follows that the infimum of the action over all transition paths (i.e., the quasipotential barrier $S$ separating the two subchimeras) vanishes. 
In \cref{fig:6}(a), we show that the quasipotential barrier does indeed vanish by applying a single perturbation $\bm{\Delta}_{sw}(\delta)$ to the coherent ring, with $\delta$ ranging from $10^{-5}$ to $10^{-15}$. 
The distribution of the number of times a transition path is found through this procedure shows that the landscape is highly nontrivial for paths of small action: Transition barriers of all heights exist, and the height distribution follows a power law.
This claim is further supported by Fig. 6(b), where we show the action for $1000$ different transition paths, each obtained by applying $\bm{\Delta}_{sw}(\delta)$ at a different time $t$ (the same initial condition is used for all simulations). 
One can see that the landscape varies wildly and the associated action spans many decades. 
As we include more transition paths, deeper and deeper valleys can be found, bringing the smallest action ever closer to zero.

\begin{figure*}[!hbt]
\centering
\includegraphics[width=1\linewidth]{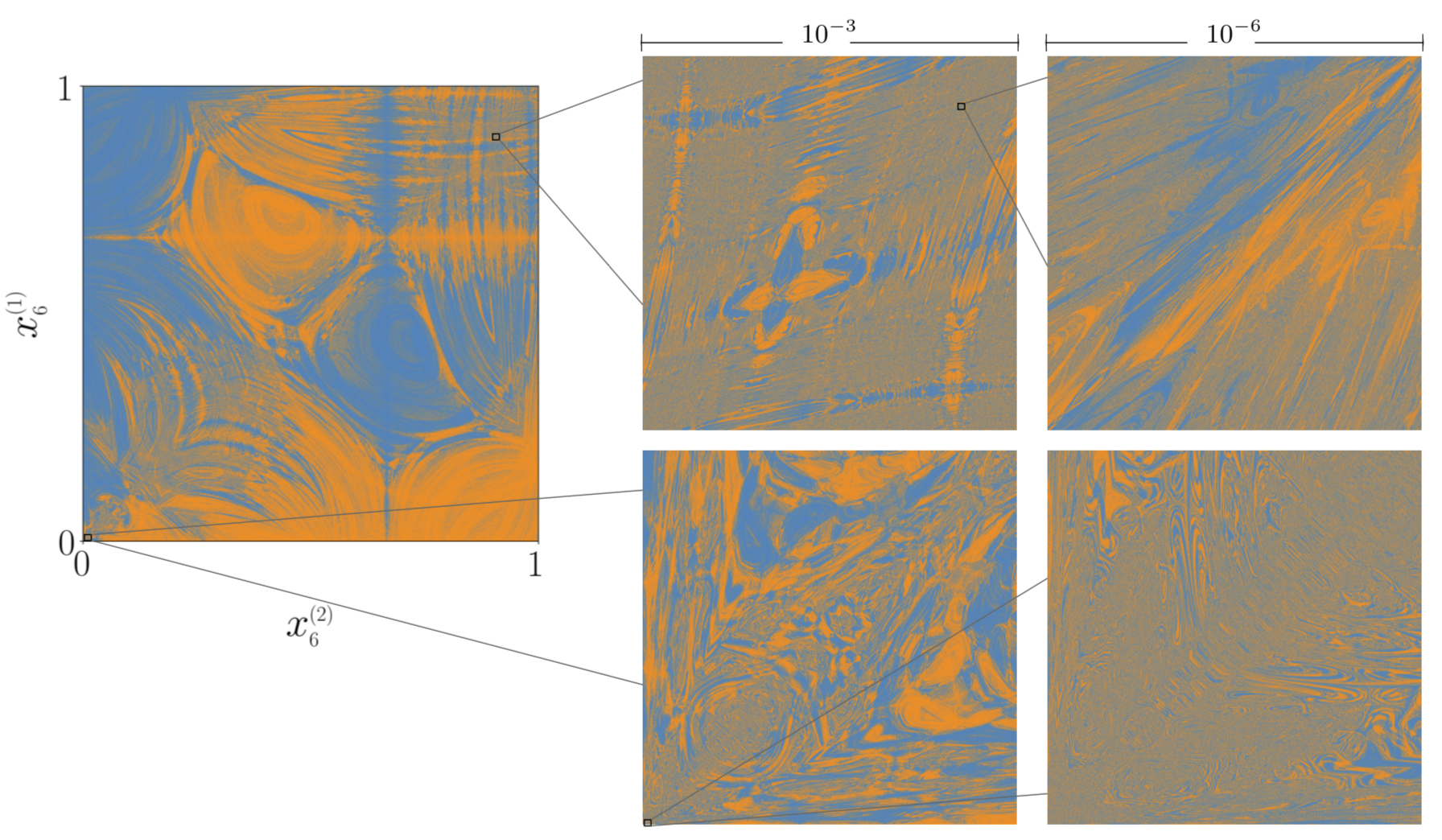} 
\vspace{-2mm}
\caption{
Two-dimensional section of the state space showing intermingled basins of the two subchimeras. 
The two basins, shown in blue and orange, are fat fractals \cite{ott2002chaos} intermingled with each other everywhere. 
Orange points are attracted to the subchimera where the first ring is synchronized, and the blue ones converge to the subchimera with the second ring synchronized. 
There is a symmetry between the two basins with respect to reflections across the diagonal, which originates from the reflection symmetry of the network. 
The areas marked for magnification are intentionally oversized to facilitate visualization.
The choice of state-space section and system parameters are specified in the text.
}
\label{fig:7}
\end{figure*}

The power-law distribution of barrier heights, in turn, gives rise to the power-law scaling of the average switching periods shown in \cref{fig:2}. 
This relationship follows because the only transition paths that matter are the ones with action comparable to the square of noise intensity.
Although there are many more higher-action paths, the probability of crossing those barriers is exponentially smaller.
The argument is further supported by the scaling exponents in \cref{fig:2,fig:6}, which differ only by a negative sign.

There are intriguing parallels between what we find here and critical phenomena in second-order phase transitions \cite{stauffer2014introduction,stanley1971phase}.
For instance, in site percolation models, the correlation (which quantifies the likelihood of two sites being connected) decays exponentially with distance when the occupation probability is $p<p_c$, but the decay changes to a power law at the critical point $p=p_c$.
Here, the average switching period scales exponentially with the inverse square of noise intensity, $\xi^{-2}$, when the quasipotential barrier has $S>0$, but it is replaced by a power law when $S=0$.
There are finite barriers of all heights between the two subchimeras when $S=0$; similarly, in percolation, there are finite clusters of all sizes at the critical point $p=p_c$.
The power laws uncovered here, however, are more robust than those from the percolation theory. 
The latter happens only at the critical point and requires fine-tuning, whereas here the power-law switching persists for a wide range of parameters.
In this sense, the analogy is perhaps closer with self-organized criticality \cite{bak1987self,bak1988self,diaz1994dynamic}, in which scale-invariance emerges in the absence of fine-tuning.

\subsection{Intermingled basins}
\label{sec:riddled}

By now, we have explained the ``anomalous'' power-law switching behavior from a first-exit model in log-error space (Sec.~\ref{sec:log space}) as well as by characterizing the action landscape of transition paths (Sec.~\ref{sec:quasi-potential}).
In those characterizations, one can catch glimpses of chaos lurking in the background, but its exact role is still unclear.
In this section, we establish a direct connection between power-law switching and riddled basins \cite{alexander1992riddled,ott1993scaling,ott1994blowout,heagy1994experimental,ott1994transition,ashwin1994bubbling,maistrenko1998transverse,aguirre2009fractal,santos2018riddling}, which is possible only for chaotic attractors \cite{ashwin1996attractor}, thus bringing the fundamental importance of the chaotic dynamics to the forefront.

Chaos has long been known to produce power laws by generating fractal structures in state space \cite{ott2002chaos}. 
For example, in the presence of fractal basin boundaries, a small uncertainty $\varepsilon$ in the initial conditions translates to an uncertainty of $A\varepsilon^\alpha$ percent on the final states, where prefactor $A$ is a constant and $\alpha$ is the uncertainty exponent given by the difference between the state-space dimension and the box-counting dimension of the basin boundary \cite{mcdonald1985fractal}.
In the case of riddled basins, the entire basin is its own (fractal) boundary and $\alpha=0$. 
This means that, for any $\varepsilon$, the $\varepsilon$-neighborhood of an arbitrary point in a riddled basin will always include points that are in the basin of some other attractor \cite{ott2002chaos}.

\begin{figure*}[!hbt] 
\centering
\includegraphics[width=1\linewidth]{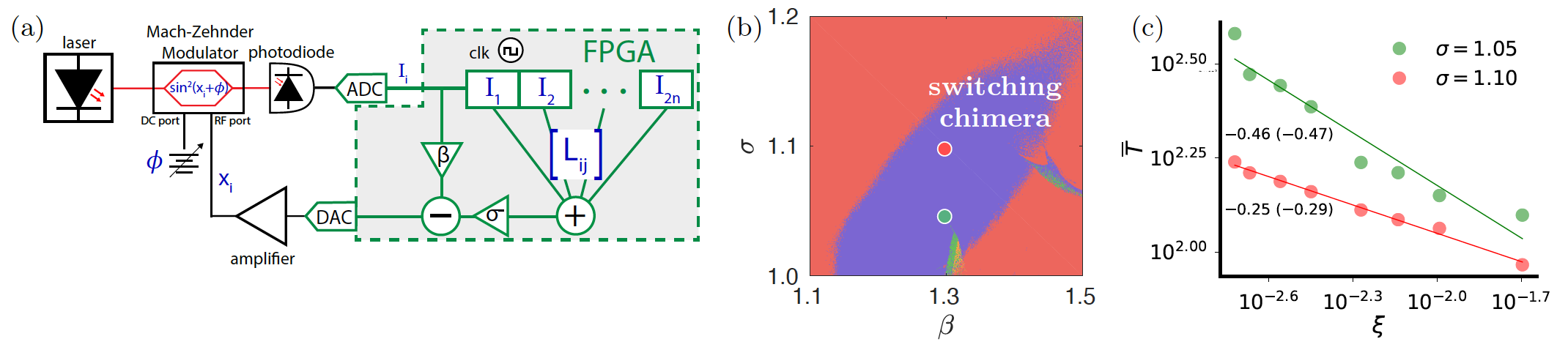}
\vspace{-4mm}
\caption{Experimental realization of globally attractive switching chimeras.
(a) Schematic diagram of the optoelectronic system, where the dashed box depicts our implementation of the coupling scheme.
(b) Parameter space color coded according to direct simulations of \cref{eq:3}. The regions shown include switching chimeras (purple), nonswitching chimeras (green), chimera death \cite{zakharova2014chimera} (yellow), and incoherence (red).
(c) Experimentally measured average switching period $\overline{T}$ as a function of the noise intensity $\xi$ for $\beta=1.3$  and two values of $\sigma$
[dots in (b)]. 
The scaling exponents annotated on the figure are obtained through linear least-square fitting applied to the relationship between $\log(\overline{T})$ and $\log(\xi)$.
The exponents obtained from experiments are in good agreement with those predicted from simulations (shown in parentheses).}
\label{fig:8}
\end{figure*}

In \cref{fig:7}, we show a two-dimensional section of the twelve-dimensional state space to visually illustrate that the attraction basin of each subchimera is riddled. 
Because the two basins are mutually riddled, they are referred to as {\it intermingled basins}. In this figure, the initial conditions for $x_6^{(1)}$ and $x_6^{(2)}$ are sampled independently from the interval $[0,1]$, while the initial conditions for the other oscillators are specified as $x_i^{(1)} = x_6^{(1)}/2$ and $x_i^{(2)} = x_6^{(2)}/2$, where $1\leq i \leq 5$.
We then simulate \cref{eq:0} for $\sigma = 1.7$ and $r = 2.95$ in the absence of noise and record the subchimera attractor each trajectory is attracted to. 
(There is nothing special about the choice of the parameters or the section of the state space, since other choices lead to similar results.) 
One can observe intricate fractal-like structures in all parts of the two-dimensional section, for all resolutions considered (up to pixels of size $10^{-10}\times10^{-10}$).
There is also a symmetry between the two basins.
If an initial condition is in the basin of one subchimera, then its mirror image reflected along the diagonal line must be in the basin of the other subchimera [i.e., if $(x^{(1)}_6,x^{(2)}_6) = (a,b)$ is blue, then $(x^{(1)}_6,x^{(2)}_6) = (b,a)$ is orange]. 
This is the result of the reflection symmetry between the two rings in \cref{fig:1}(a).

Because the basins are intermingled, the basin of one subchimera has points arbitrarily close to the other subchimera attractor, and vice versa, which gives rise to arbitrarily small transition barriers in \cref{fig:6}. 
Thus, the subchimeras are attractors in the sense of Milnor \cite{milnor1985concept} (i.e., attracts initial conditions of nonzero measure) but not in the sense of attracting an open neighborhood of initial conditions containing the attractor.

Apart from the Freidlin-Wentzell action, the perturbation magnitude $\delta$ in \cref{fig:6} can also be interpreted as a distance from the closest subchimera attractor.
The probability $p$ then measures the fraction of the state space that converges to the opposite subchimera when at distance $\delta$ from the subchimera attractor.
As the initial conditions are taken further away from one subchimera, it becomes more likely for the system to land in the basin of the other subchimera.
Conversely, as $\delta\rightarrow0$, the probability of escaping to the opposite subchimera approaches zero algebraically.
This property is visualized using a transverse section of the intermingled basins that directly connects the two subchimera attractors, as shown in \cref{sec:transveral section}.

Although arbitrarily small perturbations can drive the system out of a subchimera attractor, both subchimeras are transversally stable according to linear stability analysis. 
While seemingly incompatible, these two conditions can coexist when an attractor is transversally stable for the natural measure but unstable for some other invariant ergodic measure. 
In fact, transversal stability for the natural measure and instability for at least one other invariant ergodic measure are necessary conditions for riddled basins to occur \cite{ashwin1996attractor}.
This mathematical statement is, in its core, similar to the intuitive explanation given in Sec.~\ref{sec:power-law} on why a system can be driven away from a linearly stable state by arbitrarily small perturbations.

\section{Experimental observation of switching chimeras}
\label{sec:experiment}

Thus far, we have focused on the theoretical analysis of networks of logistic maps, which reveals remarkable features of a new chimera state, including intermingled basins and switching triggered by arbitrarily small noise. 
To demonstrate that the theoretical results can be observed under realistic conditions and for different oscillator dynamics, we perform experiments on networks of coupled optoelectronic oscillators. As we show next, our experiments confirm the existence of switching chimeras in physical systems.

\begin{figure*}[!hbt]
\centering
\includegraphics[width=1\linewidth]{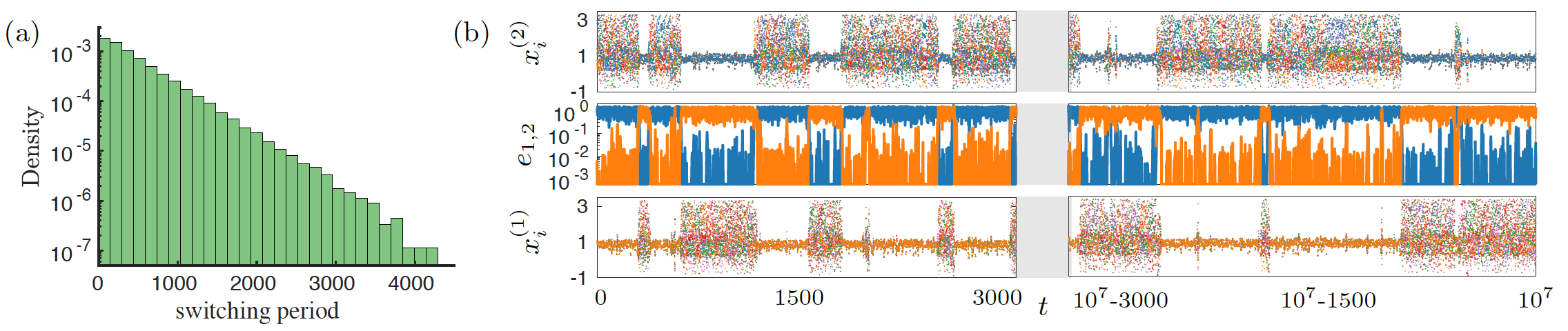}
\vspace{-4mm}
\caption{
Statistics and dynamics of a switching chimera in the experiments.
(a) Distribution of switching periods for $\beta=1.3$ and $\sigma=1.05$ [green dot in \cref{fig:8}(b)].
(b) Portion of the experimentally measured time series used to generate (a). These measurements are performed at the base-noise level of the system, which is estimated to be $0.0019$.
}
\label{fig:9}
\end{figure*}

The experimental setup is schematically shown in \cref{fig:8}(a). 
A single optoelectronic oscillator draws nonlinearity from a Mach-Zehnder modulator, which takes voltage $x$ as an input and outputs light of intensity $\sin^2(x+\phi)$. 
The operation point $\phi$ is fixed at $\pi/4$ throughout the experiments.
Time multiplexing and delays are used to realize multiple oscillators from a single time-delayed feedback loop, which reduces apparatus costs and allows for the realization of a large number of truly identical oscillators.
The oscillators are coupled together by a digital filter implemented electronically on a field-programmable gate array (FPGA) according to a predetermined Laplacian matrix $\bm{L}=\{L_{ij}\}$.
In this case, $\bm{L}$ describes the two-cluster network shown in \cref{fig:1}(a).
Further details of the optoelectronic system can be found in Refs.~\cite{hart2017experiments,hart2019delayed}.

The main source of intrinsic noise comes from the measurement of  light intensity, including the noise introduced by the analog-to-digital converter (ADC) due to its finite resolution.
To best model the experimental system, we introduce independent Gaussian noise to the oscillators at each iteration: $I\big(x_i^{(1,2)}[t]\big) = \sin^2\big(x_i^{(1,2)}[t]+\phi\big) + \xi N_i^{(1,2)}[t]$.
The dynamical equation describing the optoelectronic oscillator network can then be written as
\begin{equation}
  \begin{split}
    x_i^{(1,2)}[t+1] = & \,\beta\,I\big(x_i^{(1,2)}[t]\big) \\
    + & \sigma \left( I\big(x_{i-1}^{(1,2)}[t]\big) + I\big(x_{i+1}^{(1,2)}[t]\big) - 2I\big(x_i^{(1,2)}[t]\big) \right) \\ 
    + & c\sigma \sum_{j=1}^n \left( I\big(x_{j}^{(2,1)}[t]\big) - I\big(x_i^{(1,2)}[t]\big) \right),
\end{split}
\label{eq:3}
\end{equation}
where the noise term is implicitly included in $I$. In our experiments, we again set $c = 0.2$ and $n = 6$. 

We first sweep the parameter space of feedback strength $\beta$ and coupling strength $\sigma$ using direct simulations of \cref{eq:3}.
As shown in \cref{fig:8}(b), switching chimeras are predicted to occupy a significant portion of this space.
Inside the switching chimera region (purple), the red and green dots denote the parameters to be systematically investigated in the experiments.

The dynamics exhibited by the experimental system is in many ways qualitatively similar to that of coupled logistic maps. 
In particular, a clear pattern of irregular switching between two subchimeras is observed for suitable parameters, as shown in \cref{fig:9}(b). 
To characterize the experimental dynamics quantitatively, we first test whether the power-law relationship between the average switching time $\overline{T}$ and noise intensity $\xi$ holds in the experimental data. 
An important step in the data analysis is to estimate the level of the intrinsic experimental noise, which we do by simulating \cref{eq:3} under different $\xi$ to extract $\overline{T}$ for a range of noise intensities. 
The simulation results are then compared with the $\overline{T}$ observed in the experiments.  For both parameter sets ($\beta=1.3$, $\sigma=1.05$ and $\beta=1.3$, $\sigma=1.1$), the simulations with noise intensity $0.0019$ agree best with the experiments. 
We thus choose Gaussian noise of intensity $\xi_1$ to approximate the base-noise level intrinsic to the experimental system.
It is worth noting that this technique can, in principle, be extended to estimate the level of intrinsic noise in other oscillators, even when the noise is extremely small---an outstanding problem for which, to the best of our knowledge, no general approach currently exists.

To implement variable noise in the experiments, we introduce an additional Gaussian noise term of tunable intensity $\xi_2$ via the FPGA.
Assuming that the intrinsic and external noise terms are independent, the experimental system is effectively subject to a 
Gaussian noise of intensity $\xi = \sqrt{\xi_1^2+\xi_2^2}$.
\Cref{fig:8}(c) summarizes the experimentally measured $\overline{T}$ for different $\xi$ from the lower bound $0.0019$ all the way to $0.02$. 
Each data point is averaged over at least $20000$ experimentally observed switching events.
It can be seen that the power-law relationship holds under realistic noise levels and is robust against the imperfections typical of an experimental system.
In addition, we also perform systematic simulations to further confirm that the power-law scaling persists in the presence of a small amount of heterogeneity among the oscillators (\cref{sec:heterogeneity}).

\Cref{fig:9}(a) shows the distribution of the switching periods $T$ extracted from $45000$ switching events, for data collected from multiple experimental runs with $\beta=1.3$, $\sigma=1.05$, and $\xi_2=0$.
The distribution of periods is clearly exponential.
This is a consequence of the fact that, although the evolution of the synchronization errors $e_1$ and $e_2$ is non-Markovian (Sec.~\ref{sec:log space}), the switching events themselves are described by a Poisson process. 
In particular, the experimental data show that the waiting period until the next switching event is independent of the previous switching events.
For such a memoryless process with a constant transition rate, the time between switching events is guaranteed to be exponentially distributed \cite{roy1980first}. 

Our experimental results are further visualized using an animated spatiotemporal representation of the time-series data presented in \cref{fig:9}(b) (Supplemental Material \cite{SM}, Sec.~S2 and associated animation). 
As in the case of coupled logistic maps, the underlying state-space structure giving rise to this dynamics is the intermingled nature of the attraction basins.
Indeed, direct simulations of \cref{eq:3} confirm that the basins of the two symmetric subchimeras are intermingled (Supplemental Material \cite{SM}, Sec.~S3).

\section{Connections with biological and other physical systems}
\label{sec:connections}

A switching chimera can be seen as a chimera state whose symmetry is not broken when considering the long-term dynamics---asymptotically, one cannot distinguish between the behavior of the two clusters.
With this observation in mind, we can establish an intriguing parallel between the switching chimera and the symmetry-breaking phenomenon of dipole inversion \cite{anderson1972more}. 
Many small molecules, such as ammonia, have more than one (symmetry-broken) ground state with nonvanishing dipole moments.
However, due to quantum tunneling, an ammonia molecule switches rapidly between its two ground states, canceling out the opposite dipole moments and restoring the broken symmetry.
The same can be stated for switching chimeras, since each of the two symmetric subchimeras has a broken parity symmetry but the switching between them restores that symmetry.
For larger and heavier molecules, such as sugars or phosphorus trifluoride, dipole inversion is no longer likely to be excited by quantum tunneling or even thermal fluctuations, and, thus, the symmetry is spontaneously broken and nonvanishing dipole moments persist.
We observe that the tendency for transitioning between subchimeras also decreases in larger systems, with the average switching period growing exponentially as the number of nodes is increased (Supplemental Material \cite{SM}, Sec.~S4).

It is instructive to notice that an exponential dependence of the average switching period on system size is also observed for the magnetized states in the Ising model for any nonzero temperature below the critical point \cite{kindermann1980markov,lebowitz1999statistical}.
However, because there is a finite energy barrier to overcome for transitions between the magnetized states, the dependence of the average switching period on the inverse temperature (the analog of the inverse square of noise intensity in our systems) is not power law but instead exponential.

Switching between symmetry-broken states are not limited to physical systems. 
In particular, switching chimeras may have implications for aperiodic lateral switching in biological systems, of which interhemispheric switching in songbirds during vocal production is an example \cite{wang2008rapid}. 
Other examples of lateral switching include alternating eye movement in chameleons and fish \cite{pettigrew1999convergence}, switching in neural activity inside the two sinuses of leech hearts \cite{stewart2004networking}, and unihemispheric sleep in dolphins, birds, and other animals \cite{rattenborg2000behavioral,mathews2006asynchronous}.
A common aspect of these various processes is that they involve alternations in the activity between two approximately symmetrical lateral sides. 
Despite previous progress \cite{schmidt2008using}, the underlying mechanism of lateral switching remains elusive.
This is especially the case for aperiodic lateral switching, since such cases cannot be easily modeled by hypothesizing the existence of a central pattern generator or propagating wave dynamics, as in previous alternating chimeras \cite{ma2010robust,haugland2015self,bick2018heteroclinic}.
In the case of the songbird zebra finches, for instance, the interhemispheric switching between song-control areas of the brain is highly irregular, characterized by switching intervals ranging from $4$ to $150$ ms \cite{wang2008rapid}. 
Switching chimeras offer a simple mechanism by which a wide range of switching intervals can emerge naturally, and, thus, suggest the possibility that aperiodic lateral switching could be generated spontaneously (as opposed to, for example, being forced by neurotransmitter release \cite{lapierre2007cortical}).

\section{Concluding remarks}
\label{sec:discussions}

The theoretical, computational, and experimental results presented here offer a comprehensive characterization of a novel class of chimera states that are globally attractive and exhibit power-law switching dynamics.
We extended the Freidlin-Wentzell theory to derive the observed power-law scaling, and we demonstrated that there is no finite quasipotential barrier separating the two symmetric subchimeras.
This unexpected scaling behavior, which should be contrasted with the exponential scaling observed for typical noise-induced transitions \cite{hanggi1986escape,bolhuis2002transition}, was confirmed under realistic conditions in our experiments using networks of optoelectronic oscillators.
We also established a connection between switching chimeras and intermingled basins, which provides insight into both phenomena. 
In particular, the latter explains why switching between subchimeras occurs for arbitrarily small noise despite each subchimera being linearly stable.

We expect switching chimeras to be a common phenomenon in multilayer networks with symmetry. 
These networks are generalizations of the two-layer networks considered in Ref.~\cite{abrams2008solvable}.
In particular, switching between symmetric subchimeras is expected to be possible for networks formed by any number of identically coupled identical layers, where the layers themselves can have an arbitrary structure. 
Thus, while we focused on networks with two subchimeras, our analysis extends naturally to other states and to a larger number of switching configurations. 
From the dynamical perspective, we point to the following conditions for the emergence of power-law switching behavior: 
(i) There are two or more attractors and they are embedded in manifolds of dimension lower than that of the state space; 
(ii) each attractor is chaotic and has transversally unstable periodic orbits embedded within.
If the transitions are not restricted to chimera states, the requirement on the network structure can be further relaxed, as these conditions are often satisfied even by single-layer oscillator networks.

Finally, we note that the observed high noise sensitivity of the switching dynamics has far-reaching implications.
It can be exploited, for instance, to detect small intrinsic noise in oscillator systems---e.g., by using time multiplexing to create a network of such systems that exhibits power-law switching.
It also offers a potential explanation for irregular switching noticed in biological systems, which suggests that the dynamical behavior described here may be observed in naturally evolved processes.

\section*{Acknowledgments}

The authors thank Daniel J.\ Case for insightful discussions.
This work was supported by ARO Grant No.\ W911NF-19-1-0383 and ONR Grant No.\ N000141612481.

\appendix

\section{Linear stability analysis of chimera states}
\label{sec:stability}

In order to assess the linear stability of a chimera state, we calculate the synchronization stability in the coherent cluster while taking into account the influence of the incoherent cluster.
This calculation can be done efficiently using a generalization of the master stability function formalism developed in Ref.~\cite{hart2019topological}, which is tailored to describe the synchronization stability of individual clusters.

Consider a network of $2n$ diffusively coupled identical oscillators:
\begin{equation}
    x_i[t+1] = f(x_i[t]) - \sigma \sum_{j=1}^{2n} L_{ij} h(x_j[t]), 
\end{equation}
where $x_i$ is the state of the $i$-th oscillator, $f$ is the mapping function governing the uncoupled dynamics of each oscillator, 
$\bm{L} = \{L_{ij}\}$ is the Laplacian matrix describing the structure of an undirected network with two nonintertwined identical clusters, $h$ is the interaction function, and $\sigma>0$ is the coupling strength.

Let $\widetilde{\bm{L}}$ be the $n\times n$ Laplacian matrix that encodes the intracluster connection inside the coherent cluster, $\mu$ be the total strength of intercluster connections each oscillator in the coherent cluster receives from the incoherent cluster, and $x_1=x_2=\dots=x_n=s$ be the synchronization manifold for the $n$ oscillators in the coherent cluster.
The variational equation describing the evolution of the deviation away from $s$ can be written as
\begin{equation}
  \delta\bm{X}[t+1] = \left( \mathds{1}_{n} \otimes f'(s[t]) - \sigma \widehat{\bm{L}} \otimes h'(s[t]) \right) \delta\bm{X}[t],
  \label{eq:s2}
\end{equation}
where $\mathds{1}_n$ is the identity matrix, $\widehat{\bm{L}} = \widetilde{\bm{L}} + \mu\mathds{1}_n$, $\delta\bm{X} = (\delta x_1,\dots,\delta x_n)^\intercal = (x_1 - s,\dots,x_n - s)^\intercal$, and $ \otimes$ denotes the Kronecker product.
Although the incoherent cluster does not enter the equation explicitly, it influences the matrix $\widehat{\bm{L}}$ and the synchronization trajectory $s[t]$ through the intercluster coupling.
We note that the input from the incoherent cluster faithfully accounts for the state of those oscillators and is time dependent in general.

\Cref{eq:s2} can be decoupled into $n$ independent equations by diagonalizing $\widehat{\bm{L}}$:
\begin{equation}
    \eta_i[t+1] = \big( f'(s[t]) - \sigma \widehat{v}_i h'(s[t]) \big) \eta_i[t],
    \label{eq:s3}
\end{equation}
where $\bm{\eta} = (\eta_1,\dots,\eta_n)^\intercal$ is $\delta\bm{X}$ expressed in the new coordinates that diagonalize $\widehat{\bm{L}}$ and $\widehat{v}_i = \widetilde{v}_i + \mu$ are the eigenvalues of $\widehat{\bm{L}}$ in ascending order.
Synchronization in the coherent cluster is stable if and only if $\Lambda(\sigma \widehat{v}_i)<0$ for $i=2,\dots, n$, where 
\begin{equation}
	\Lambda(\sigma \widehat{v}_i) = \lim_{T\rightarrow\infty}\frac{1}{T}\sum_{t=0}^{T-1}\ln\Big\rvert f'(s[t]) - \sigma \widehat{v}_i h'(s[t]) \Big\rvert
\end{equation}
is the Lyapunov exponent of \cref{eq:s3} and $\widehat{v}_2, \dots, \widehat{v}_n$ represent the perturbation modes transverse to the synchronization manifold of the coherent cluster.
The maximum transverse Lyapunov exponent (MTLE) determining the synchronization stability is $\max_{2\leq i \leq n} \Lambda (\sigma \widehat{v}_i )$.
A chimera state is stable for $\xi=0$ if the MTLE for synchronization in the coherent cluster is negative under the influence of the incoherent cluster.

\section{Dominant switching route}
\label{sec:swb route}

Here, we provide more evidence that short-wavelength bifurcation is the dominant mechanism to initiate switching between the two symmetric subchimeras. 
Again, we simulate \cref{eq:0} to extract the average switching period $\overline{T}$ for various levels of noise intensity $\xi$, but this time the short-wavelength component $\bm{\Delta}_{sw}$ is filtered out from the noise applied to each ring.
If a short-wavelength bifurcation is indeed the dominant route for switching, then one would expect the average switching period to become independent of the noise intensity after filtration.

\begin{figure}[tb]
\centering
\subfloat[]{
\includegraphics[width=.95\columnwidth]{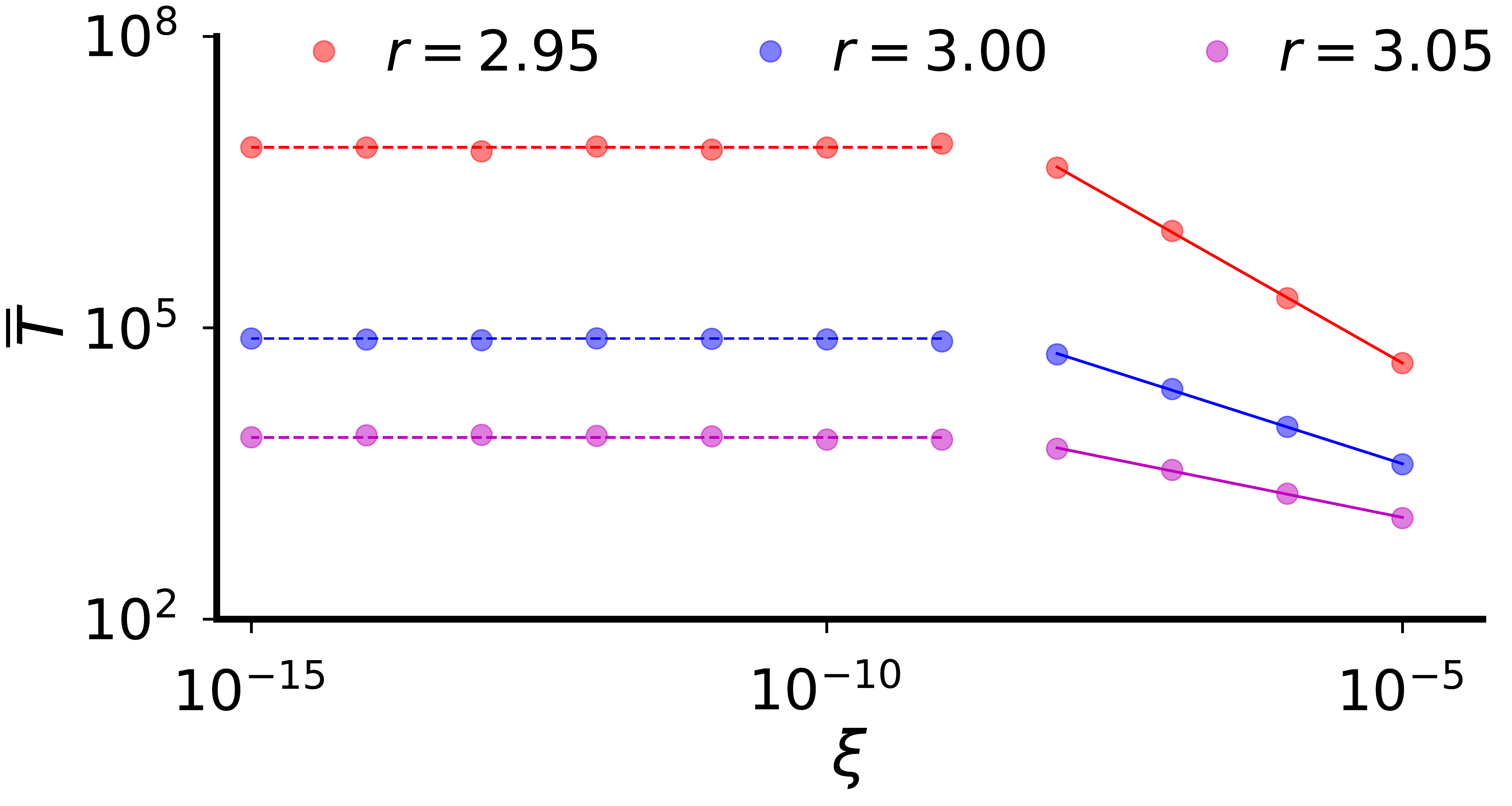}
}
\vspace{-6mm}
\caption{Average switching period $\overline{T}$ as a function of noise intensity $\xi$ for various $r$. The system is the network of logistic maps in \cref{fig:1}(a) for $\sigma=1.7$, and the noise is Gaussian (but with the short-wavelength component filtered out). The flatness of the fitting lines below $\xi=10^{-9}$ confirms that short-wavelength bifurcation is the dominant route for chimera switching.}
\label{fig:s6}
\end{figure}

This is exactly the case shown in \cref{fig:s6}, where the slope becomes completely flat for each $r$ when the noise intensity goes below $10^{-9}$ (compare with \cref{fig:2}).
Due to the presence of round-off errors in our simulations, whose short-wavelength component cannot be filtered, switching can still be observed in the flat region at a rate induced by the round-off errors (noise intensity around $10^{-16}$).
When the noise intensity goes above $10^{-9}$, new switching pathways besides the short-wavelength bifurcation start to become available, as demonstrated by the resulting decrease of the average switching period.

\begin{figure}[t]
\centering
\subfloat[]{
\includegraphics[width=\columnwidth]{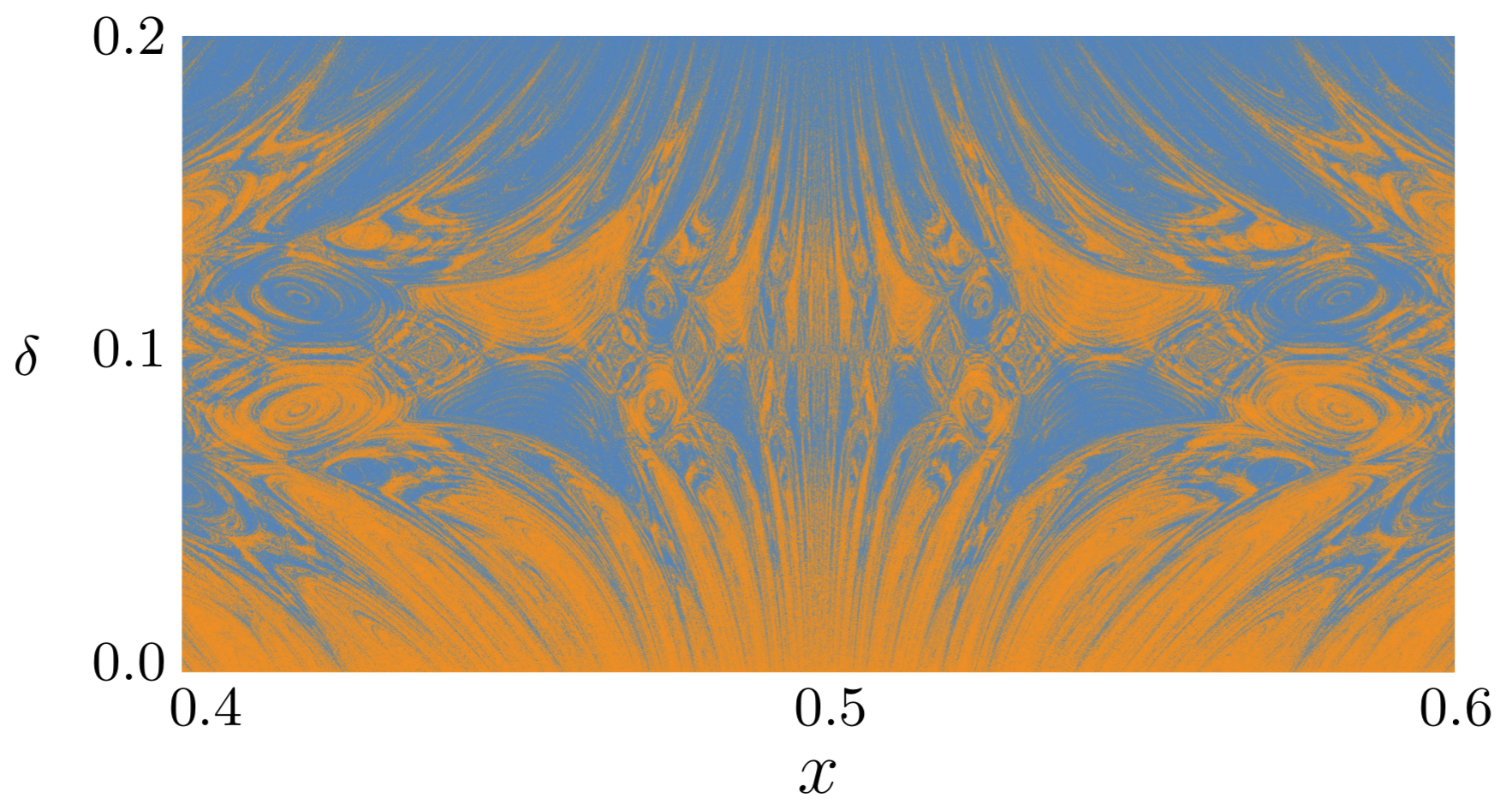}
}
\vspace{-5mm}
\caption{Transversal section of the intermingled basins that directly connects the two symmetric subchimeras. This corresponds to a different state-space section of the system considered in \cref{fig:7}.}
\label{fig:s7}
\end{figure}

\begin{figure}[!tb]
\centering
\subfloat[]{
\includegraphics[width=.9\columnwidth]{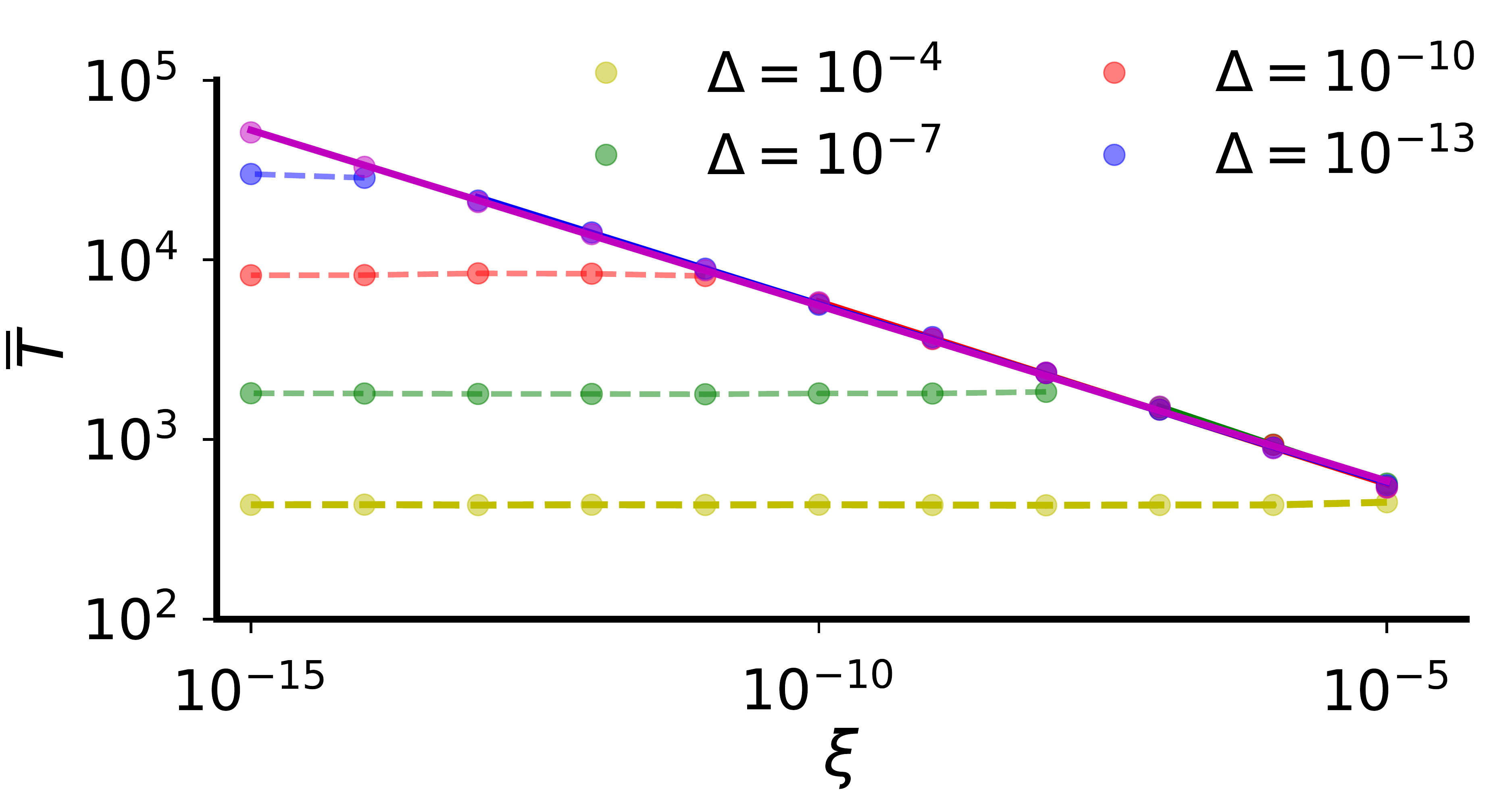}
}
\vspace{-5mm}
\caption{Effect of oscillator heterogeneity on the switching behavior determined from direct simulations. The solid line indicates the power-law scaling for $\xi \geq \Delta$, which is precisely the scaling observed in the absence of oscillator heterogeneity. 
For each of the four levels of heterogeneity $\Delta$ considered, when $\xi < \Delta$ the effect of heterogeneity becomes dominant and the average switching period $\overline{T}$ becomes independent of $\xi$.}
\label{fig:s11}
\end{figure}

\section{Transversal section of intermingled basins}
\label{sec:transveral section}

\Cref{fig:s7} shows the intermingled basins for a two-dimensional section of the state space for the logistic map system described by \cref{eq:0}.
This section is defined by
\begin{equation}
	\bm{x}^{(1)} = x\mathds{1}_6 + \bm{\Delta}_{sw}(\delta), \quad \bm{x}^{(2)} = x\mathds{1}_6 + \bm{\Delta}_{sw}(\delta_{\max}-\delta),
\end{equation}
where $\delta_{\max}$ is taken to be $0.2$. 
For $\delta = 0$, the first ring is synchronized and the second ring is incoherent (orange subchimera), while for $\delta = \delta_{\max} $, the second ring is synchronized and the first ring is incoherent (blue subchimera).
Thus, this section of the state space directly connects the two symmetric subchimeras.
As one approaches the orange (blue) subchimera, the points become predominantly orange (blue), but no matter how close $\delta$ is to zero ($\delta_{\max}$), speckles of blue (orange) dots can always be found.

\vspace{8mm}

\section{Robustness against oscillator heterogeneity}
\label{sec:heterogeneity}

In \cref{fig:s11}, we quantify the effect of oscillator heterogeneity on the switching dynamics, explicitly demonstrating the robustness of the switching chimeras. 
We start from a system of identical oscillators (the system in \cref{fig:1} for $r=3$ and $\sigma=1.7$) and introduce independent random perturbations to the parameter $r$ of each oscillator, drawn from a Gaussian distribution of zero mean and standard deviation $\Delta$.

For $\xi \geq \Delta$, the average switching periods in the homogeneous and heterogeneous systems become indistinguishable, with both following a well-defined power-law distribution on noise intensity. 
For $\xi < \Delta$, the effect of heterogeneity dominates the effect of noise; as a result, the average switching period (dashed lines) branches out of the original power-law relation (solid line) and approaches a constant determined by $\Delta$. 
These results are largely independent of the particular realization of oscillator heterogeneity. 

\bibliographystyle{prx_ref}
\bibliography{net_dyn}

\end{document}